\documentclass[runningheads]{llncs}
\usepackage{graphicx}
\usepackage{caption}
\usepackage{xcolor}
\usepackage{soul}
\usepackage{hyperref}

\begin{document}
%
\title{Deep Learning Framework for Real-time Fetal Brain Segmentation in MRI}
%
%
\author{Razieh Faghihpirayesh\inst{1,2} \and
Davood Karimi\inst{2} \and
Deniz Erdo{\u{g}}mu\c{s}\inst{1} \and
Ali Gholipour\inst{2}}

\authorrunning{R. Faghihpirayesh et al.}
%

\institute{Electrical and Computer Engineering Department, Northeastern University
\and
Radiology Department, Boston Children’s Hospital; and Harvard Medical School\\
Boston, MA, 02115, USA\\
\email{raziehfaghih@ece.neu.edu}}

\maketitle              
%

\begin{abstract}
Fetal brain segmentation is an important first step for slice-level motion correction and slice-to-volume reconstruction in fetal MRI.
Fast and accurate segmentation of the fetal brain on fetal MRI is required to achieve real-time fetal head pose estimation and motion tracking for slice re-acquisition and steering. To address this critical unmet need, in this work 
we analyzed the speed-accuracy performance of a variety of deep neural network models, and devised a symbolically small convolutional neural network that combines spatial details at high resolution with context features extracted at lower resolutions. We used multiple branches with skip connections to maintain high accuracy while devising a parallel combination of convolution and pooling operations as an input downsampling module to further reduce inference time.
We trained our model as well as eight alternative, state-of-the-art networks with manually-labeled fetal brain MRI slices and tested on two sets of normal and challenging
test cases.
Experimental results show
that our network achieved the highest accuracy and lowest inference time among all of the compared state-of-the-art real-time segmentation methods. We achieved average Dice scores of 97.99\% and 84.04\% on the normal and challenging test sets, respectively, with an inference time of 3.36 milliseconds per image on an NVIDIA GeForce RTX 2080 Ti.
Code, data, and the trained models are available at \href{https://github.com/bchimagine/real_time_fetal_brain_segmentation}{this repo}.

\keywords{Fetal MRI \and Fetal brain \and Real-time segmentation.}

\end{abstract}

\section{Introduction}
\label{section:intro}

Fetal MRI is an important tool for diagnosis of abnormalities of the fetal brain during pregnancy due to its superior soft tissue contrast compared to ultrasound. However, MRI is very susceptible to motion and fetuses can move significantly during MRI scans. To mitigate this problem, fast MRI acquisition techniques are used to obtain stacks of 2D slices. Super-resolution techniques can then reconstruct 3D images from these 2D slices~\cite{gholipour2010robust,kuklisova2012reconstruction,kainz2015fast,ebner2018automated,ebner2020automated,uus2020deformable}. Segmentation of the brain in slices can improve inter-slice motion correction and super-resolution reconstruction~\cite{wang2020uncertainty}. Real-time fetal brain segmentation on slices is needed to enable real-time fetal head pose estimation, motion tracking, and slice navigation~\cite{salehi2018pose}.

While several studies have addressed 3D fetal brain segmentation on stack-of-slices or reconstructed fetal MRI scans~\cite{anquez2009automatic,salehi2017auto,khalili2017automatic,ebner2018automated}, only a few studies have addressed the more challenging task of segmenting the fetal brain on every slice. Keraudren et al.~\cite{keraudren2014automated} developed a method based on support vector machines and random forests. More recent works have almost exclusively been based on deep learning (DL), and in particular convolutional neural networks (CNNs). These methods are more suitable for real-time applications because they can harness the parallel computation capabilities of Graphical Processing Units (GPUs)~\cite{rampun2019automated}. Salehi et al.~\cite{salehi2018real} used a DL method based on the U-Net architecture~\cite{ronneberger2015u}. Wang et al.~\cite{wang2019aleatoric} computed aleatoric uncertainty and used test time augmentation to improve the accuracy of fetal brain segmentation on 2D slices. While these works focused on improving segmentation accuracy, none of them addressed the accuracy-speed trade-off. To address this gap, in this paper we focused on improving inference speed as well as accuracy. 

Many applications demand real-time image processing. This demand has given rise to a growing body of real-time DL-based methods~\cite{papadeas2021real}. The majority of these works have aimed at reducing the computation time by devising lighter or specialized network architectures. A typical example of architectural innovations is the depthwise-separable convolution, which breaks down a 3D convolution operation into a succession of 2D and 1D convolutions~\cite{howard2017mobilenets}. Another approach to reducing the computational cost is channel shuffling as used in ShuffleNets~\cite{zhang2018shufflenet,ma2018shufflenet}. Gamal et al.~\cite{gamal2018shuffleseg} proposed ShuffleSeg based on ShuffleNet by using ShuffleNet with grouped convolutions, channel shuffling as encoder, and FCN8s~\cite{long2015fully} as decoder. ENet~\cite{paszke2016enet} uses early downsampling of the input to extract relevant image features while reducing the image size. ENet also uses a much smaller decoder module than in typical symmetric encoder–decoder architectures~\cite{badrinarayanan2017segnet,ronneberger2015u}.

Two-branch networks~\cite{poudel2018contextnet,yu2018bisenet} are another way to design faster models and are among the fastest existing methods. Unlike standard models where the entire network learns low-level and high-level details, in two-branch networks these two tasks are performed by two separate branches. A shallow branch captures spatial details and generates high-resolution feature representation, while a deeper but lightweight branch learns high-level semantic context. ContextNet~\cite{poudel2018contextnet}, Fast-SCNN~\cite{poudel2019fast}, and BiseNet~\cite{yu2018bisenet} are examples of two-branch networks. An important consideration in designing these architectures is to ensure proper integration of high-level and low-level context information. For example, ICNet~\cite{zhao2018icnet} computes a multi-resolution set of feature maps and employs a cascade feature fusion unit to fuse these feature maps, whereas DFANet~\cite{li2019dfanet} uses several interconnected encoding paths to add high-level context into the encoded features. 

In this work we aimed to design a network with proper, efficient integration of high-level and low-level information to achieve high accuracy and very fast inference in fetal brain MRI segmentation. To achieve this, we developed a new, efficient CNN-based network, which we term Real-time Fetal Brain Segmentation Network (RFBSNet). RFBSNet uses an encoder-decoder architecture with forward connections to retain accuracy; and a two-branch architecture with an input downsampling module to achieve fast inference. We compared RFBSNet with eight alternative state-of-the-art DL models. In the following sections, we provide a detailed description of our methods, data, results, and analysis.


\section{Materials and Methods}

\label{section:method}

\subsection{Proposed Network Architecture}
\begin{figure}[t!]
    \includegraphics[width=0.95\textwidth]{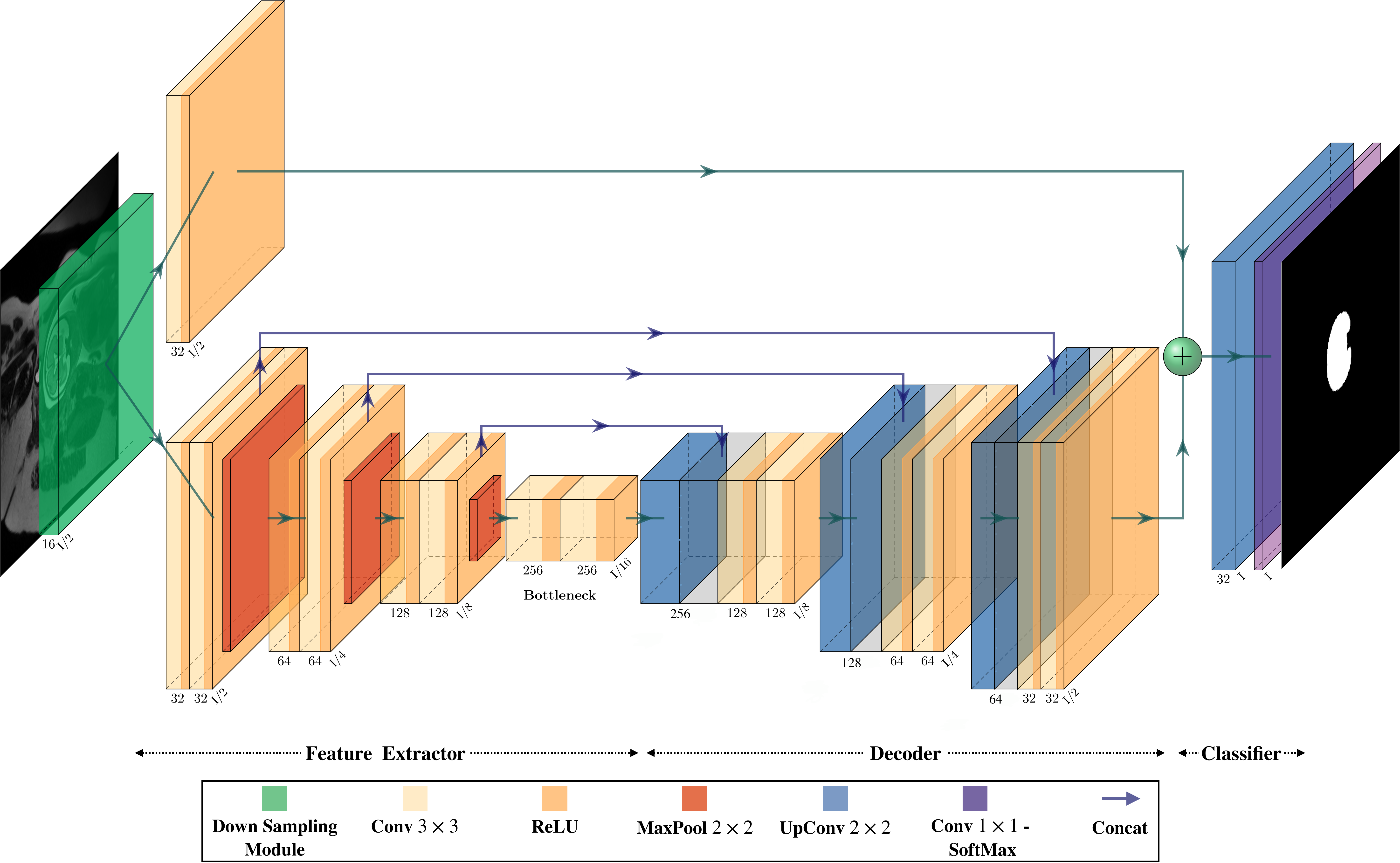}
    \caption{Overview of the proposed architecture (RFBSNet). It consists of an input down-sampling module, a feature extractor, a decoder, and a classifier, with two branches and forward connections from the feature extractor to the decoder. All modules are built using classical convolution layers using operations shown in the figure legend. The detail of the down sampling module is shown in Fig.~\ref{fig:init}. Numbers next to each block show the number of channels, while the length indicates the spatial size considering the input size of I.}
    \label{fig:RFBSNet}
\end{figure}

We designed RFBSNet to strike a balance between inference speed and accuracy. Fig.~\ref{fig:RFBSNet} shows the layout of RFBSNet. It contains an input downsampling module, a feature extractor, a decoder, and a classification module. In the following, we describe each structure module in more detail.

\subsubsection{Input downsampling module.}

\begin{figure}[t!]
    \centering
    \includegraphics[width=0.35\textwidth]{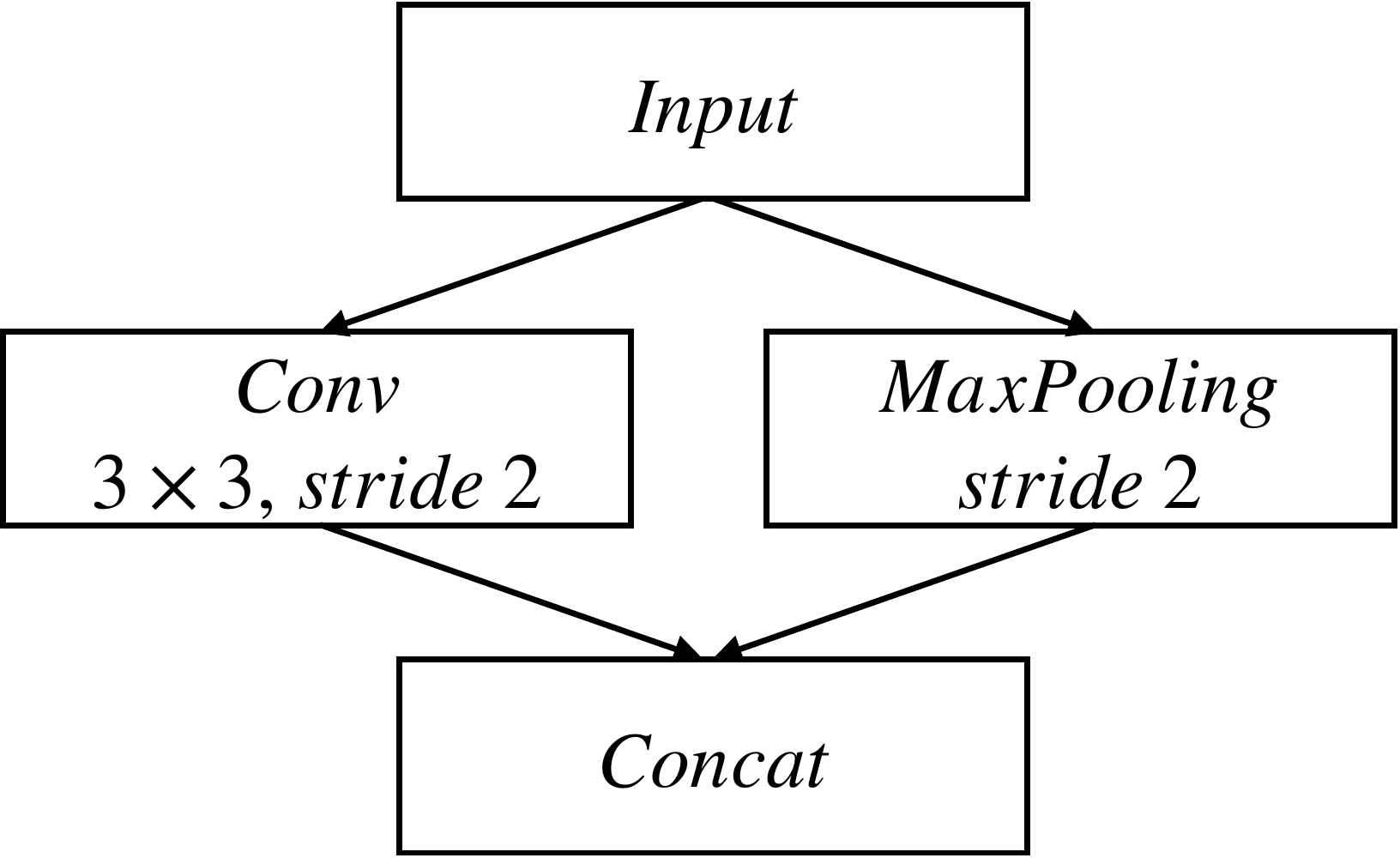}
    \caption{The input downsampling module in RFBSNet consists of a convolution and a max pooling path. The outputs are concatenated to build a feature map.} \label{fig:init}
\end{figure}

The first module in our proposed network is a downsampling module that reduces the size of the input image while also providing high resolution spatial information into the classifier module using a forward path. Input downsampling can greatly speed up the network by significantly reducing the amount of computation performed by all down-stream network layers. When this down-sampling is not excessive and is carried out using learnable functions, such as a convolution layer, the loss in segmentation accuracy can be very small. However, excessive down-sampling can result in a loss of important detail such as fine object boundaries~\cite{paszke2016enet}. Besides, downsampling the input image by a factor of $m$ would require upsampling by the same factor in order to obtain a segmentation map with the same size as the input image. Although upsampling can also be accomplished using learnable transposed convolutions, it can result in further loss of fine detail if it is excessive. To avoid these negative effects, we used a down-sampling module, shown in Fig.~\ref{fig:init}, that consists of two paths: (1) a max-pooling path with non-overlapping $2 \times 2$ windows, and (2) a convolutional layer with $3 \times 3$ kernels. The outputs of these two paths are concatenated.

\subsubsection{Feature Extractor.}

This module is responsible for learning multi-resolution image features for accurate segmentation.  Our feature extractor module shares its computation of the first few layers with the input downsampling module. This parameter sharing not only reduces the computational complexity of the network, it also improves the segmentation accuracy. 
In this architecture, we used one convolutional layer followed by ReLU in the shallow branch to encode detailed spatial information.
The deep feature extractor branch of RFBSNet provides sufficient receptive field.
We deployed U-Net style~\cite{ronneberger2015u} forward skip connections to fuse multi-resolution features into the decoder module. 

\subsubsection{Decoder and Classification modules.}

A U-Net-type decoder with skip connections upsamples the features learned by the different sections of the feature extractor module to the size of the feature maps generated by the input downsampling module. These feature maps are finally fused together using a simple addition. 
In the classification module, an upsampling layer and a pointwise convolution layer are applied to the fused feature maps. A softmax operation is applied to the final layer to generate class probability maps.



\subsection{Alternative Methods and Evaluation Metrics}

We compare the proposed RFBSNet to two standard networks for medical image segmentation (U-Net~\cite{ronneberger2015u} and SegNet~\cite{badrinarayanan2017segnet}), as well as several recent architectures that have been proposed for real-time segmentation (see Table~\ref{table:results}). We also introduce ShuffleSeg V2 following the design of ShuffleSeg~\cite{gamal2018shuffleseg}. It employs ShuffleNet V2~\cite{ma2018shufflenet} as encoder and FCN8s~\cite{long2015fully} as decoder.

The accuracy of all methods are evaluated and compared using the Dice similarity coefficient and Intersection-over-Union IoU, also known as Jaccard index, metrics defined as $Dice(P, R) = \frac{2|P\cap R|}{|P|+|R|} = \frac{2TP}{2TP+FP+FN}$ and $IoU(P,R) = \frac{|P\cap R|}{|P\cup R|} = \frac{TP}{TP+FP+FN}$ respectively. where P is predicted brain mask, R is a ground truth mask and TP, FP, and FN are the true positive, false positive, and false negative rates, respectively.
We assess segmentation speed in terms of the average inference time and standard deviation with 100 iterations for each method while using batch size of 1. In addition, we report the number of floating point operations (FLOPs) and the number of trainable parameters for each network.


 \subsection{Data, Implementation, and Training}
The fetal MRI data used in this study were acquired using 3T Siemens scanners. The study was approved by the institutional review board; and written informed consent was obtained from all research MRI participants. For each subject, multiple half-Fourier single shot turbo spin echo (HASTE) images were acquired with in-plane resolution of 1 to 1.25 mm, and slice thickness of 2 to 4 mm. The gestational ages of the fetuses at the time of scans were between 22 to 38 weeks (mean=29, stdev=5).
In total, 3496 2D fetal MRI slices (of 131 stacks from 23 fetal MRI sessions) were included in the training and validation procedure (80\% train, 20\% validation). A set of 840 2D slices (17 stacks) of two normal fetuses without severe artifacts was used as normal test set, and a set of 136 2D slices of a fetal MRI scan with artifacts (from 4 stacks) was used as the challenging test set. An experienced annotator carefully segmented the fetal brain in every slice of all these stacks. We used these manual segmentations as the ground truth for model training and evaluation.

 All experiments were conducted with an NVIDIA GeForce RTX 2080 Ti, using TensorFlow and Keras 2.6.0. All models were trained with a batch size of 8 and input image size of $256 \times 256$. We used Dice similarity coefficient between the network predictions and the ground truth as the training loss function. The learning rate for each of the compared networks was tuned separately. For our model we used an initial learning rate of $1\times 10^{-4}$, which we multiplied by 0.9 after every 2000 training steps. We trained each model for 100 epochs using Adam optimization~\cite{kingma2014adam} of stochastic gradient descent.
\section{Results}

\begin{table}[t!]
    \caption{Comparing RFBSNet with eight state-of-the-art methods based on average Dice (aDice) and average IOU (aIoU) on normal and challenging test sets. This table also shows the number of FLOPS (in giga FLOPS), the number of training parameters, and the inference time (in ms) of each method. RFBSNet achieved the highest accuracy and the best inference speed among all methods.}\label{table:results}
    \centering
    \resizebox{\textwidth}{!}{%
    \begin{tabular}{l|c|c|c|c|c|c|c}
        \hline
         & \multicolumn{2}{c|}{Normal} & \multicolumn{2}{c|}{Challenging} & FLOPs & \#Trainable& Inference Time \\
        \cline{2-5}
        Model & aDice (\%) & aIoU (\%) & aDice (\%) & aIoU (\%) & (G) & Parameters & (ms) \\
        \hline
        ICNet~\cite{zhao2018icnet} & 85.43 & 77.00 & 65.15 & 58.00 & 1.81 & 6,710,914 & 12.91$\pm$0.39\\
        ENet~\cite{paszke2016enet} & 95.42 & 91.54 & 78.05 & 69.00 & 0.975 & 362,838 & 20.63$\pm$0.05\\
        Fast-SCNN~\cite{poudel2019fast} & 86.87 & 78.81 & 69.34 & 60.80 & 0.517 & 1,593,222 & 5.74$\pm$0.09\\
        DFANet~\cite{li2019dfanet} & 78.59 & 69.19 & 65.47 & 57.97 & 0.248 & 418,354 & 22.90$\pm$0.32\\
        ShuffleSeg~\cite{gamal2018shuffleseg} & 91.34 & 85.05 & 79.38 & 70.35 & 0.374 & 940,722 & 12.05$\pm$1.11\\
        ShuffleSeg V2~\cite{ma2018shufflenet}& 89.97 & 83.05 & 77.36 & 68.26 & 1.15 & 3,043,294 & 8.64$\pm$0.25\\
        SegNet~\cite{badrinarayanan2017segnet} & 96.17 & 92.85 & 88.95 & 81.74 & 79.9 & 29,441,986 & 10.16$\pm$0.18\\
        U-Net~\cite{ronneberger2015u} & 97.93 & 96.02 & 85.81 & 77.63 & 102 & 34,512,258 & 9.82$\pm$0.04\\
        \textbf{RFBSNet} & \textbf{97.99} & \textbf{96.12} & \textbf{84.04} & \textbf{75.50} & \textbf{5.32} & \textbf{2,154,328} & \textbf{3.36$\pm$0.02}\\
         
        \hline
    \end{tabular}%
    }
\end{table}

\label{section:exp_result}

Table~\ref{table:results} summarizes the performance of our proposed RFBSNet compared to other methods.
In terms of almost all evaluation criteria,
RFBSNet outperformed the standard methods as
well as other state-of-the-art real-time segmentation models. It reached 97.99\% aDice (average Dice of all test images), 96.12\% aIoU on normal and 86.04\% aDice, 75.50\% aIoU on challenging test sets with outstanding inference time of 3.36 ms. Indeed, our network can run on a single GPU in real time, i.e., it runs as soon as a single MRI slice is acquired and reconstructed. We note that RFBSNet performed better than the standard medical image segmentation network U-Net in terms of both accuracy and speed while having $\approx 14$ times less number of parameters and FLOPs. Our method also outperformed other real-time segmentation methods in both accuracy and speed while representing comparable number of parameters and FLOPs.

We performed paired t-tests with a $p$ value threshold of 0.001 to test if the segmentation accuracy, in terms of Dice and IoU on the test sets, for our model was higher than other models. These tests showed that our model was significantly more accurate than all competing real-time segmentation models on the normal and challenging test images in terms of both Dice and IoU. Our model was also significantly more accurate than SegNet. However, the differences with the U-Net were not statistically significant ($p \approx 0.3 $). We note that in addition to computation time, both UNet and SegNet require high GPU memory which may limit their use with larger images on standard GPUs.

\begin{figure}[t!]
    \centering
    \includegraphics[width=0.85\textwidth]{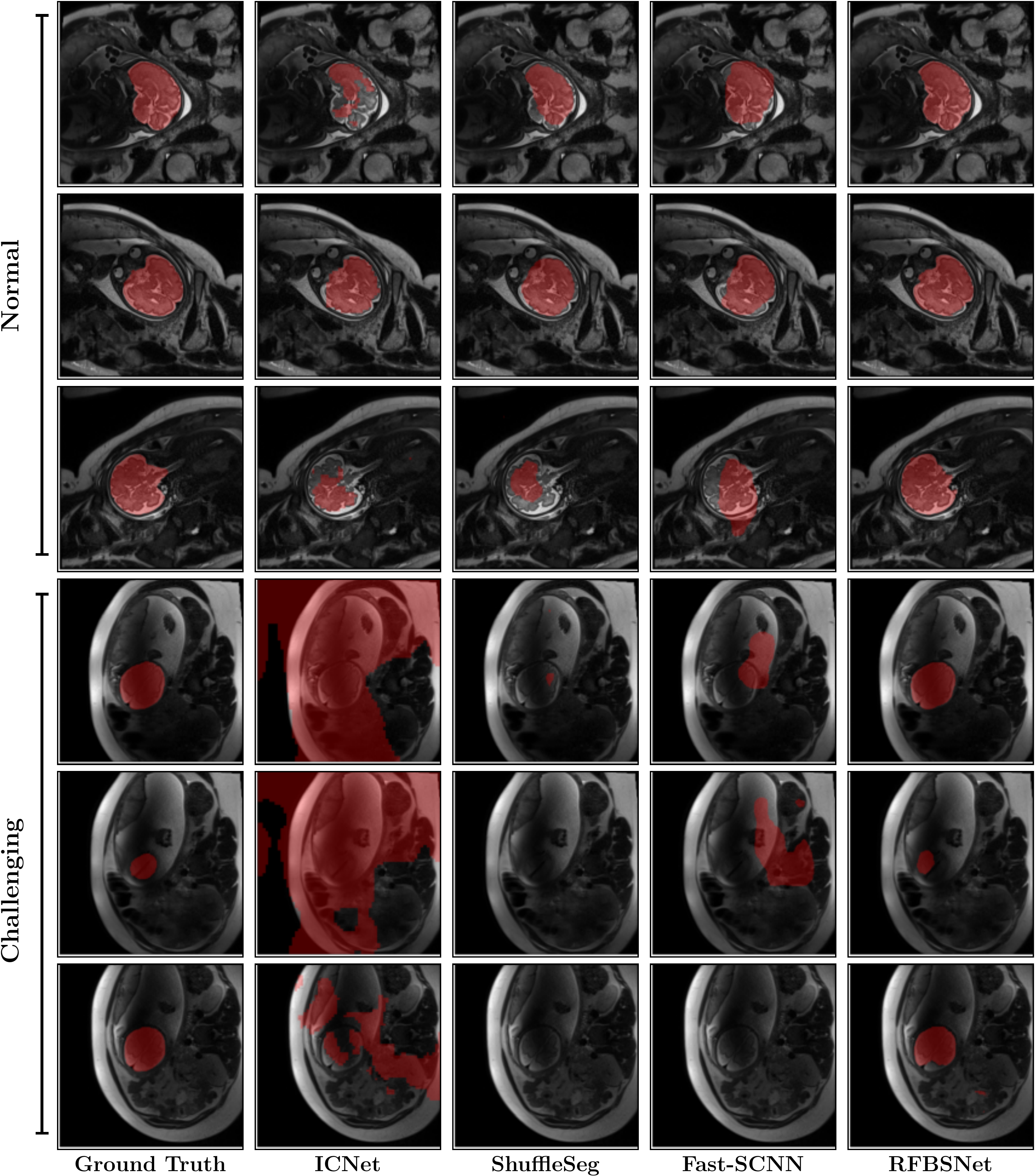}
    \caption{Representative examples of predicted brain masks overlaid on original fetal MRI slices for normal cases (top 3 rows) and challenging cases (bottom 3 rows). Note that RFBSNet correctly segmented the brain in all slices of this challenging test case which was not segmented by the other methods.} \label{fig:mask}
\end{figure}

Example segmentation results can be seen in Fig.~\ref{fig:mask}. In addition to our own method, in this figure we have shown the results of those competing methods that were proposed originally for real-time segmentation. As these representative examples show, compared to those other methods, which showed large segmentation errors and often completely failed to segment the fetal brain on challenging images, our method accurately segmented both normal and challenging images. 

\section{Analysis and Discussion}

\label{section:discussion}
%

\begin{figure}[t!]
    \centering
    \captionsetup{width=0.45\textwidth}
    \begin{minipage}[t]{.5\textwidth}
      \centering
      \includegraphics[width=0.95\linewidth]{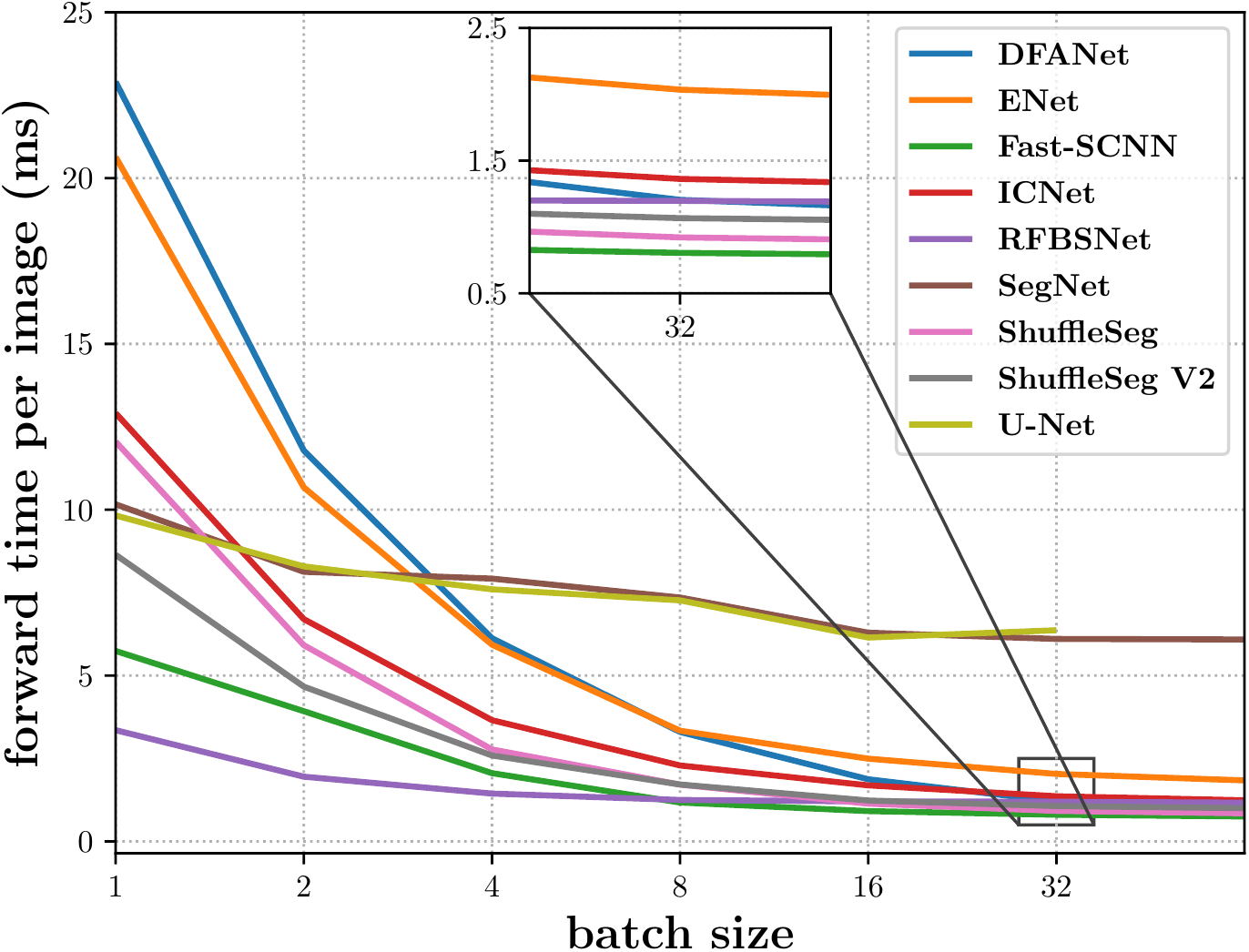}
      \captionof{figure}{Inference time vs. batch size. Missing data points are due to lack of enough system memory required to process larger batches. For batch size of 1 (needed in real-time application), RFBSNet performed best.}
      \label{fig:timevsbs}
    \end{minipage}%
    \begin{minipage}[t]{.5\textwidth}
      \centering
      \includegraphics[width=0.95\linewidth]{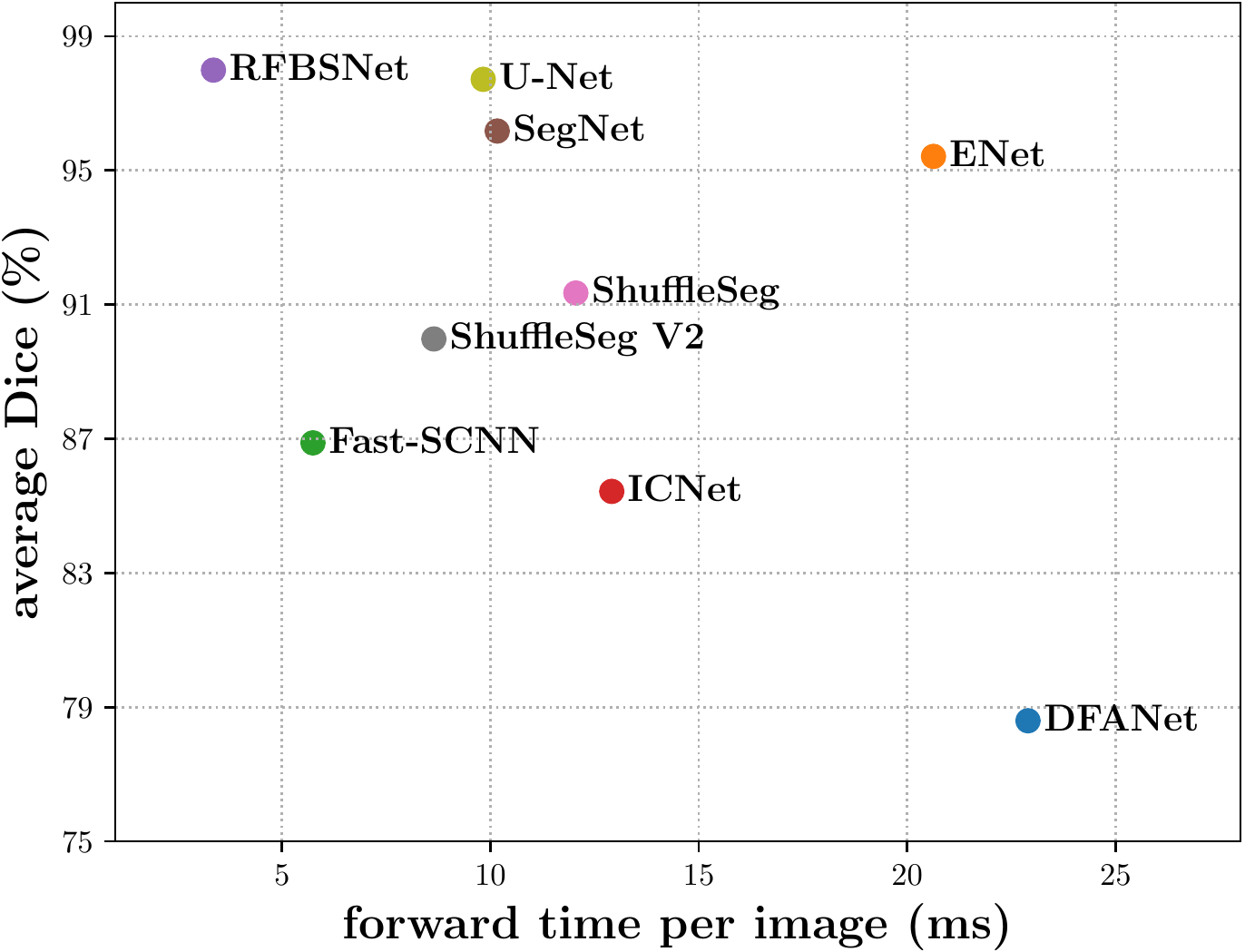}
      \captionof{figure}{Inference time vs. average Dice coefficient. This chart shows the speed-accuracy trade-off comparison on the test set at batch size of 1 for all methods that were implemented and compared in this study.}
      \label{fig:timevsdice}
    \end{minipage}
\end{figure}

We further analyzed the performance of the networks and the speed-accuracy balance. Fig.~\ref{fig:timevsbs} shows the run-time of different networks, which were implemented and compared in this study, as a function of batch size. The main observation from this figure is the dramatic shift in the order of networks as the batch size decreases. For batch sizes larger than four, real-time segmentation networks were significantly faster than standard networks. However, the order began to reverse as the batch size decreased to 1. At a batch size of 1, the processing time (per image) for all real-time segmentation networks is several times longer than those of the batch size of 10. In fact, four out of the six competing real-time segmentation networks became slower than U-Net and SegNet. Our proposed network, on the other hand, is comparable with other real-time networks for large batches and it is the fastest of all networks for a batch size of 1 (which is used for real-time inference). Fig.~\ref{fig:timevsdice} shows the speed-accuracy trade-off comparison of all methods on the normal test set with the top left corner being the optimal performance. 

By design, our proposed network (RFBSNet) achieved high accuracy and very fast inference at batch size of 1 for real-time image segmentation. The learnable input downsampling module in RFBSNet helped reduce the computations while providing a capacity to learn full spatial image resolution details. This module resembled the shallow spatial path~\cite{poudel2019fast} of two-branch models~\cite{poudel2018contextnet}. Our feature extractor module, on the other hand, can be compared to the deep low-resolution branch of those prior works. This module helped ensure a high segmentation accuracy.

As explained in section~\ref{section:intro}, depthwise-separable convolutions (DWSConv)~\cite{howard2017mobilenets} are a common design choice for reducing the computational cost of DL models. However, because DWSConv involves far fewer floating point operations than standard 2D convolutions, its execution time on a GPU is dominated by the memory access latency \cite{lu2021optimizing}. To overcome this bottleneck, existing implementations of DWSConv try to accelerate execution by using large batch sizes. However, this strategy does not work in applications where inference is highly latency-sensitive and when smaller batch sizes have to be used. Hence, DWSConvs does not result in fast models in a real-time application such as fetal brain segmentation, where a test-time batch size of one is desired.

\section{Conclusion}

\label{section:conclusion}

In this paper, we proposed a fast and accurate CNN based network for fetal brain segmentation in MRI.
Our design combines spatial details at high resolution with context features extracted at lower resolutions. We also used multiple branches with skip connections to maintain high accuracy while devising a parallel combination of convolution and pooling operations as an input down-sampling module to further reduce inference time. Experimental results showed the superiority of our proposed network compared to standard and state-of-the-art real-time segmentation models. We also demonstrated the effect of batch size at the time of inference on the latency. With an inference time of $<5$ ms, our model can segment the fetal brain in real time, leaving sufficient time for the rest of the processing that is needed for real-time motion analysis and slice navigation. 


\section{Acknowledgements}
This study was supported in part by the National Institutes of Health (NIH) under award numbers R01EB031849, R01NS106030, and R01EB032366; and in part by the Office of the Director of the NIH under award number S10OD0250111. The content of this paper is solely the responsibility of the authors and does not necessarily represent the official views of the NIH.

\clearpage
\bibliographystyle{splncs04}
\bibliography{ref}

\begin{thebibliography}{10}
\providecommand{\url}[1]{\texttt{#1}}
\providecommand{\urlprefix}{URL }
\providecommand{\doi}[1]{https://doi.org/#1}

\bibitem{anquez2009automatic}
Anquez, J., Angelini, E.D., Bloch, I.: Automatic segmentation of head
  structures on fetal mri. In: 2009 IEEE International Symposium on Biomedical
  Imaging: From Nano to Macro. pp. 109--112. IEEE (2009)

\bibitem{badrinarayanan2017segnet}
Badrinarayanan, V., Kendall, A., Cipolla, R.: Segnet: A deep convolutional
  encoder-decoder architecture for image segmentation. IEEE transactions on
  pattern analysis and machine intelligence  \textbf{39}(12),  2481--2495
  (2017)

\bibitem{ebner2018automated}
Ebner, M., Wang, G., Li, W., Aertsen, M., Patel, P.A., Aughwane, R., Melbourne,
  A., Doel, T., David, A.L., Deprest, J., et~al.: An automated localization,
  segmentation and reconstruction framework for fetal brain mri. In:
  International Conference on Medical Image Computing and Computer-Assisted
  Intervention. pp. 313--320. Springer (2018)

\bibitem{ebner2020automated}
Ebner, M., Wang, G., Li, W., Aertsen, M., Patel, P.A., Aughwane, R., Melbourne,
  A., Doel, T., Dymarkowski, S., De~Coppi, P., et~al.: An automated framework
  for localization, segmentation and super-resolution reconstruction of fetal
  brain mri. NeuroImage  \textbf{206},  116324 (2020)

\bibitem{gamal2018shuffleseg}
Gamal, M., Siam, M., Abdel-Razek, M.: Shuffleseg: Real-time semantic
  segmentation network. arXiv preprint arXiv:1803.03816  (2018)

\bibitem{gholipour2010robust}
Gholipour, A., Estroff, J.A., Warfield, S.K.: Robust super-resolution volume
  reconstruction from slice acquisitions: application to fetal brain mri. IEEE
  transactions on medical imaging  \textbf{29}(10),  1739--1758 (2010)

\bibitem{howard2017mobilenets}
Howard, A.G., Zhu, M., Chen, B., Kalenichenko, D., Wang, W., Weyand, T.,
  Andreetto, M., Adam, H.: Mobilenets: Efficient convolutional neural networks
  for mobile vision applications. arXiv preprint arXiv:1704.04861  (2017)

\bibitem{kainz2015fast}
Kainz, B., Steinberger, M., Wein, W., Kuklisova-Murgasova, M., Malamateniou,
  C., Keraudren, K., Torsney-Weir, T., Rutherford, M., Aljabar, P., Hajnal,
  J.V., et~al.: Fast volume reconstruction from motion corrupted stacks of 2d
  slices. IEEE transactions on medical imaging  \textbf{34}(9),  1901--1913
  (2015)

\bibitem{keraudren2014automated}
Keraudren, K., Kuklisova-Murgasova, M., Kyriakopoulou, V., Malamateniou, C.,
  Rutherford, M.A., Kainz, B., Hajnal, J.V., Rueckert, D.: Automated fetal
  brain segmentation from 2d mri slices for motion correction. NeuroImage
  \textbf{101},  633--643 (2014)

\bibitem{khalili2017automatic}
Khalili, N., Moeskops, P., Claessens, N.H., Scherpenzeel, S., Turk, E., Heus,
  R.d., Benders, M.J., Viergever, M.A., Pluim, J.P., I{\v{s}}gum, I.: Automatic
  segmentation of the intracranial volume in fetal mr images. In: Fetal, Infant
  and Ophthalmic Medical Image Analysis, pp. 42--51. Springer (2017)

\bibitem{kingma2014adam}
Kingma, D.P., Ba, J.: Adam: A method for stochastic optimization. arXiv
  preprint arXiv:1412.6980  (2014)

\bibitem{kuklisova2012reconstruction}
Kuklisova-Murgasova, M., Quaghebeur, G., Rutherford, M.A., Hajnal, J.V.,
  Schnabel, J.A.: Reconstruction of fetal brain mri with intensity matching and
  complete outlier removal. Medical image analysis  \textbf{16}(8),  1550--1564
  (2012)

\bibitem{li2019dfanet}
Li, H., Xiong, P., Fan, H., Sun, J.: Dfanet: Deep feature aggregation for
  real-time semantic segmentation. In: Proceedings of the IEEE/CVF Conference
  on Computer Vision and Pattern Recognition. pp. 9522--9531 (2019)

\bibitem{long2015fully}
Long, J., Shelhamer, E., Darrell, T.: Fully convolutional networks for semantic
  segmentation. In: Proceedings of the IEEE conference on computer vision and
  pattern recognition. pp. 3431--3440 (2015)

\bibitem{lu2021optimizing}
Lu, G., Zhang, W., Wang, Z.: Optimizing depthwise separable convolution
  operations on gpus. IEEE Transactions on Parallel and Distributed Systems
  \textbf{33}(1),  70--87 (2021)

\bibitem{ma2018shufflenet}
Ma, N., Zhang, X., Zheng, H.T., Sun, J.: Shufflenet v2: Practical guidelines
  for efficient cnn architecture design. In: Proceedings of the European
  conference on computer vision (ECCV). pp. 116--131 (2018)

\bibitem{papadeas2021real}
Papadeas, I., Tsochatzidis, L., Amanatiadis, A., Pratikakis, I.: Real-time
  semantic image segmentation with deep learning for autonomous driving: A
  survey. Applied Sciences  \textbf{11}(19), ~8802 (2021)

\bibitem{paszke2016enet}
Paszke, A., Chaurasia, A., Kim, S., Culurciello, E.: Enet: A deep neural
  network architecture for real-time semantic segmentation. arXiv preprint
  arXiv:1606.02147  (2016)

\bibitem{poudel2018contextnet}
Poudel, R.P., Bonde, U., Liwicki, S., Zach, C.: Contextnet: Exploring context
  and detail for semantic segmentation in real-time. arXiv preprint
  arXiv:1805.04554  (2018)

\bibitem{poudel2019fast}
Poudel, R.P., Liwicki, S., Cipolla, R.: Fast-scnn: Fast semantic segmentation
  network. In: British Machine Vision Conference (2019)

\bibitem{rampun2019automated}
Rampun, A., Jarvis, D., Griffiths, P., Armitage, P.: Automated 2d fetal brain
  segmentation of mr images using a deep u-net. In: Asian Conference on Pattern
  Recognition. pp. 373--386. Springer (2019)

\bibitem{ronneberger2015u}
Ronneberger, O., Fischer, P., Brox, T.: U-net: Convolutional networks for
  biomedical image segmentation. In: International Conference on Medical image
  computing and computer-assisted intervention. pp. 234--241. Springer (2015)

\bibitem{salehi2017auto}
Salehi, S.S.M., Erdogmus, D., Gholipour, A.: Auto-context convolutional neural
  network (auto-net) for brain extraction in magnetic resonance imaging. IEEE
  transactions on medical imaging  \textbf{36}(11),  2319--2330 (2017)

\bibitem{salehi2018real}
Salehi, S.S.M., Hashemi, S.R., Velasco-Annis, C., Ouaalam, A., Estroff, J.A.,
  Erdogmus, D., Warfield, S.K., Gholipour, A.: Real-time automatic fetal brain
  extraction in fetal mri by deep learning. In: 2018 IEEE 15th International
  Symposium on Biomedical Imaging (ISBI 2018). pp. 720--724. IEEE (2018)

\bibitem{salehi2018pose}
Salehi, S.S.M., Khan, S., Erdogmus, D., Gholipour, A.: Real-time deep pose
  estimation with geodesic loss for image-to-template rigid registration. IEEE
  transactions on medical imaging  \textbf{38}(2),  470--481 (2018)

\bibitem{uus2020deformable}
Uus, A., Zhang, T., Jackson, L.H., Roberts, T.A., Rutherford, M.A., Hajnal,
  J.V., Deprez, M.: Deformable slice-to-volume registration for motion
  correction of fetal body and placenta mri. IEEE transactions on medical
  imaging  \textbf{39}(9),  2750--2759 (2020)

\bibitem{wang2020uncertainty}
Wang, G., Aertsen, M., Deprest, J., Ourselin, S., Vercauteren, T., Zhang, S.:
  Uncertainty-guided efficient interactive refinement of fetal brain
  segmentation from stacks of mri slices. In: International Conference on
  Medical Image Computing and Computer-Assisted Intervention. pp. 279--288.
  Springer (2020)

\bibitem{wang2019aleatoric}
Wang, G., Li, W., Aertsen, M., Deprest, J., Ourselin, S., Vercauteren, T.:
  Aleatoric uncertainty estimation with test-time augmentation for medical
  image segmentation with convolutional neural networks. Neurocomputing
  \textbf{338},  34--45 (2019)

\bibitem{yu2018bisenet}
Yu, C., Wang, J., Peng, C., Gao, C., Yu, G., Sang, N.: Bisenet: Bilateral
  segmentation network for real-time semantic segmentation. In: Proceedings of
  the European conference on computer vision (ECCV). pp. 325--341 (2018)

\bibitem{zhang2018shufflenet}
Zhang, X., Zhou, X., Lin, M., Sun, J.: Shufflenet: An extremely efficient
  convolutional neural network for mobile devices. In: Proceedings of the IEEE
  conference on computer vision and pattern recognition. pp. 6848--6856 (2018)

\bibitem{zhao2018icnet}
Zhao, H., Qi, X., Shen, X., Shi, J., Jia, J.: Icnet for real-time semantic
  segmentation on high-resolution images. In: Proceedings of the European
  conference on computer vision (ECCV). pp. 405--420 (2018)

\end{thebibliography}

\end{document}